\documentclass[aps,prd,superscriptaddress,showpacs]{revtex4}

\usepackage{epsfig}
\usepackage{lscape}
\usepackage{graphicx}
\usepackage{dcolumn}
\usepackage{bm}
\usepackage{setspace}
\begin{document}
\setlength{\baselineskip}{6.0ex}
\renewcommand{\thetable}{\Roman{table}}

\title{A search for the exotic meson $X(5568)$ 
with the Collider Detector at Fermilab}
\affiliation{Institute of Physics, Academia Sinica, Taipei, Taiwan 11529, Republic of China}
\affiliation{Argonne National Laboratory, Argonne, Illinois 60439, USA}
\affiliation{University of Athens, 157 71 Athens, Greece}
\affiliation{Institut de Fisica d'Altes Energies, ICREA, Universitat Autonoma de Barcelona, E-08193, Bellaterra (Barcelona), Spain}
\affiliation{Baylor University, Waco, Texas 76798, USA}
\affiliation{Istituto Nazionale di Fisica Nucleare Bologna, \ensuremath{^{kk}}University of Bologna, I-40127 Bologna, Italy}
\affiliation{University of California, Davis, Davis, California 95616, USA}
\affiliation{University of California, Los Angeles, Los Angeles, California 90024, USA}
\affiliation{Instituto de Fisica de Cantabria, CSIC-University of Cantabria, 39005 Santander, Spain}
\affiliation{Carnegie Mellon University, Pittsburgh, Pennsylvania 15213, USA}
\affiliation{Enrico Fermi Institute, University of Chicago, Chicago, Illinois 60637, USA}
\affiliation{Comenius University, 842 48 Bratislava, Slovakia; Institute of Experimental Physics, 040 01 Kosice, Slovakia}
\affiliation{Joint Institute for Nuclear Research, RU-141980 Dubna, Russia}
\affiliation{Duke University, Durham, North Carolina 27708, USA}
\affiliation{Fermi National Accelerator Laboratory, Batavia, Illinois 60510, USA}
\affiliation{University of Florida, Gainesville, Florida 32611, USA}
\affiliation{Laboratori Nazionali di Frascati, Istituto Nazionale di Fisica Nucleare, I-00044 Frascati, Italy}
\affiliation{University of Geneva, CH-1211 Geneva 4, Switzerland}
\affiliation{Glasgow University, Glasgow G12 8QQ, United Kingdom}
\affiliation{Harvard University, Cambridge, Massachusetts 02138, USA}
\affiliation{Division of High Energy Physics, Department of Physics, University of Helsinki, FIN-00014, Helsinki, Finland; Helsinki Institute of Physics, FIN-00014, Helsinki, Finland}
\affiliation{University of Illinois, Urbana, Illinois 61801, USA}
\affiliation{The Johns Hopkins University, Baltimore, Maryland 21218, USA}
\affiliation{Institut f\"{u}r Experimentelle Kernphysik, Karlsruhe Institute of Technology, D-76131 Karlsruhe, Germany}
\affiliation{Center for High Energy Physics: Kyungpook National University, Daegu 702-701, Korea; Seoul National University, Seoul 151-742, Korea; Sungkyunkwan University, Suwon 440-746, Korea; Korea Institute of Science and Technology Information, Daejeon 305-806, Korea; Chonnam National University, Gwangju 500-757, Korea; Chonbuk National University, Jeonju 561-756, Korea; Ewha Womans University, Seoul, 120-750, Korea}
\affiliation{Ernest Orlando Lawrence Berkeley National Laboratory, Berkeley, California 94720, USA}
\affiliation{University of Liverpool, Liverpool L69 7ZE, United Kingdom}
\affiliation{University College London, London WC1E 6BT, United Kingdom}
\affiliation{Centro de Investigaciones Energeticas Medioambientales y Tecnologicas, E-28040 Madrid, Spain}
\affiliation{Massachusetts Institute of Technology, Cambridge, Massachusetts 02139, USA}
\affiliation{University of Michigan, Ann Arbor, Michigan 48109, USA}
\affiliation{Michigan State University, East Lansing, Michigan 48824, USA}
\affiliation{Institution for Theoretical and Experimental Physics, ITEP, Moscow 117259, Russia}
\affiliation{University of New Mexico, Albuquerque, New Mexico 87131, USA}
\affiliation{The Ohio State University, Columbus, Ohio 43210, USA}
\affiliation{Okayama University, Okayama 700-8530, Japan}
\affiliation{Osaka City University, Osaka 558-8585, Japan}
\affiliation{University of Oxford, Oxford OX1 3RH, United Kingdom}
\affiliation{Istituto Nazionale di Fisica Nucleare, Sezione di Padova, \ensuremath{^{ll}}University of Padova, I-35131 Padova, Italy}
\affiliation{University of Pennsylvania, Philadelphia, Pennsylvania 19104, USA}
\affiliation{Istituto Nazionale di Fisica Nucleare Pisa, \ensuremath{^{mm}}University of Pisa, \ensuremath{^{nn}}University of Siena, \ensuremath{^{oo}}Scuola Normale Superiore, I-56127 Pisa, Italy, \ensuremath{^{pp}}INFN Pavia, I-27100 Pavia, Italy, \ensuremath{^{qq}}University of Pavia, I-27100 Pavia, Italy}
\affiliation{University of Pittsburgh, Pittsburgh, Pennsylvania 15260, USA}
\affiliation{Purdue University, West Lafayette, Indiana 47907, USA}
\affiliation{University of Rochester, Rochester, New York 14627, USA}
\affiliation{The Rockefeller University, New York, New York 10065, USA}
\affiliation{Istituto Nazionale di Fisica Nucleare, Sezione di Roma 1, \ensuremath{^{rr}}Sapienza Universit\`{a} di Roma, I-00185 Roma, Italy}
\affiliation{Mitchell Institute for Fundamental Physics and Astronomy, Texas A\&M University, College Station, Texas 77843, USA}
\affiliation{Istituto Nazionale di Fisica Nucleare Trieste, \ensuremath{^{ss}}Gruppo Collegato di Udine, \ensuremath{^{tt}}University of Udine, I-33100 Udine, Italy, \ensuremath{^{uu}}University of Trieste, I-34127 Trieste, Italy}
\affiliation{University of Tsukuba, Tsukuba, Ibaraki 305, Japan}
\affiliation{Tufts University, Medford, Massachusetts 02155, USA}
\affiliation{Waseda University, Tokyo 169, Japan}
\affiliation{Wayne State University, Detroit, Michigan 48201, USA}
\affiliation{University of Wisconsin-Madison, Madison, Wisconsin 53706, USA}
\affiliation{Yale University, New Haven, Connecticut 06520, USA}

\author{T.~Aaltonen}
\affiliation{Division of High Energy Physics, Department of Physics, University of Helsinki, FIN-00014, Helsinki, Finland; Helsinki Institute of Physics, FIN-00014, Helsinki, Finland}
\author{S.~Amerio\ensuremath{^{ll}}}
\affiliation{Istituto Nazionale di Fisica Nucleare, Sezione di Padova, \ensuremath{^{ll}}University of Padova, I-35131 Padova, Italy}
\author{D.~Amidei}
\affiliation{University of Michigan, Ann Arbor, Michigan 48109, USA}
\author{A.~Anastassov\ensuremath{^{w}}}
\affiliation{Fermi National Accelerator Laboratory, Batavia, Illinois 60510, USA}
\author{A.~Annovi}
\affiliation{Laboratori Nazionali di Frascati, Istituto Nazionale di Fisica Nucleare, I-00044 Frascati, Italy}
\author{J.~Antos}
\affiliation{Comenius University, 842 48 Bratislava, Slovakia; Institute of Experimental Physics, 040 01 Kosice, Slovakia}
\author{G.~Apollinari}
\affiliation{Fermi National Accelerator Laboratory, Batavia, Illinois 60510, USA}
\author{J.A.~Appel}
\affiliation{Fermi National Accelerator Laboratory, Batavia, Illinois 60510, USA}
\author{T.~Arisawa}
\affiliation{Waseda University, Tokyo 169, Japan}
\author{A.~Artikov}
\affiliation{Joint Institute for Nuclear Research, RU-141980 Dubna, Russia}
% ...

\collaboration{CDF Collaboration}
\altaffiliation[With visitors from]{
\ensuremath{^{a}}University of British Columbia, Vancouver, BC V6T 1Z1, Canada,
\ensuremath{^{b}}Istituto Nazionale di Fisica Nucleare, Sezione di Cagliari, 09042 Monserrato (Cagliari), Italy,
\ensuremath{^{c}}University of California Irvine, Irvine, CA 92697, USA,
\ensuremath{^{d}}Institute of Physics, Academy of Sciences of the Czech Republic, 182~21, Czech Republic,
\ensuremath{^{e}}CERN, CH-1211 Geneva, Switzerland,
\ensuremath{^{f}}Cornell University, Ithaca, NY 14853, USA,
\ensuremath{^{g}}University of Cyprus, Nicosia CY-1678, Cyprus,
\ensuremath{^{h}}Office of Science, U.S. Department of Energy, Washington, DC 20585, USA,
\ensuremath{^{i}}University College Dublin, Dublin 4, Ireland,
\ensuremath{^{j}}ETH, 8092 Z\"{u}rich, Switzerland,
\ensuremath{^{k}}University of Fukui, Fukui City, Fukui Prefecture, Japan 910-0017,
\ensuremath{^{l}}Universidad Iberoamericana, Lomas de Santa Fe, M\'{e}xico, C.P. 01219, Distrito Federal,
\ensuremath{^{m}}University of Iowa, Iowa City, IA 52242, USA,
\ensuremath{^{n}}Kinki University, Higashi-Osaka City, Japan 577-8502,
\ensuremath{^{o}}Kansas State University, Manhattan, KS 66506, USA,
\ensuremath{^{p}}Brookhaven National Laboratory, Upton, NY 11973, USA,
\ensuremath{^{q}}Istituto Nazionale di Fisica Nucleare, Sezione di Lecce, Via Arnesano, I-73100 Lecce, Italy,
\ensuremath{^{r}}Queen Mary, University of London, London, E1 4NS, United Kingdom,
\ensuremath{^{s}}University of Melbourne, Victoria 3010, Australia,
\ensuremath{^{t}}Muons, Inc., Batavia, IL 60510, USA,
\ensuremath{^{u}}Nagasaki Institute of Applied Science, Nagasaki 851-0193, Japan,
\ensuremath{^{v}}National Research Nuclear University, Moscow 115409, Russia,
\ensuremath{^{w}}Northwestern University, Evanston, IL 60208, USA,
\ensuremath{^{x}}University of Notre Dame, Notre Dame, IN 46556, USA,
\ensuremath{^{y}}Universidad de Oviedo, E-33007 Oviedo, Spain,
\ensuremath{^{z}}CNRS-IN2P3, Paris, F-75205 France,
\ensuremath{^{aa}}Universidad Tecnica Federico Santa Maria, 110v Valparaiso, Chile,
\ensuremath{^{bb}}Sejong University, Seoul 143-747, Korea,
\ensuremath{^{cc}}The University of Jordan, Amman 11942, Jordan,
\ensuremath{^{dd}}Universite catholique de Louvain, 1348 Louvain-La-Neuve, Belgium,
\ensuremath{^{ee}}University of Z\"{u}rich, 8006 Z\"{u}rich, Switzerland,
\ensuremath{^{ff}}Massachusetts General Hospital, Boston, MA 02114 USA,
\ensuremath{^{gg}}Harvard Medical School, Boston, MA 02114 USA,
\ensuremath{^{hh}}Hampton University, Hampton, VA 23668, USA,
\ensuremath{^{ii}}Los Alamos National Laboratory, Los Alamos, NM 87544, USA,
\ensuremath{^{jj}}Universit\`{a} degli Studi di Napoli Federico II, I-80138 Napoli, Italy
}
\noaffiliation

\begin{abstract}
A search for the exotic meson $X(5568)$ decaying into the
$B^0_s \pi^{\pm}$ final state 
is performed using data corresponding to $9.6~\textrm{fb}^{-1}$ from
 $p{\bar p}$ collisions at 
$\sqrt{s} = 1960$ GeV recorded by the Collider Detector at Fermilab.  
No evidence for this state
is found and an upper limit of 6.7\% at the 95\% confidence level
is set on the fraction
of $B^0_s$ produced through the $X(5568) \rightarrow B^0_s \, \pi^{\pm}$ 
process.
\end{abstract}
\maketitle

The new and unexpected structure in the $B^0_s \, \pi^{\pm}$ 
final state recently observed 
by the D0 collaboration \cite{DZero} in $p{\bar p}$ collisions at 
$\sqrt{s} = 1960$ GeV
cannot be interpreted as a
meson composed of a quark-antiquark pair.
This reported signal, named $X(5568)$, is measured
with a mass of $5567.8 \pm 2.9 ^{+0.9}_{-1.9}$ MeV/$c^2$ and a width of
$21.9 \pm 6.4 ^{+5.0}_{-2.5}$ MeV/$c^2$.
Several collaborations in both electron-positron and
hadronic collision experiments have found evidence for other exotic
hadron candidates formed with
four or more quarks \cite{Olsen}. 
The $B^0_s \, \pi^{\pm}$ final state contains four quark flavors, 
which cannot result
from the decay of any standard meson.
An observation of this state, if confirmed, would represent the first
tetraquark (four-quark) candidate containing four different quark flavors. 
However, efforts by the LHCb and CMS collaborations to confirm the
$X(5568)$ provide no supporting evidence for its existence \cite{LHCb,CMS}. 

In this Letter we report the results of a search for the exotic meson $X(5568)$.
This search is made in
$p\overline{p}$ collisions 
at a center of mass energy of 1.96 TeV using the
Collider Detector at Fermilab (CDF II) by reconstructing 
the decay chain 
$X(5568)\rightarrow B^0_s\, \pi^{\pm}$,
$B^0_s \rightarrow J/\psi \, \phi$, $J/\psi \rightarrow \mu^+ \, \mu^-$,
and $\phi \rightarrow K^+ \, K^-$.
These studies use
the full CDF II data sample corresponding to an integrated luminosity of
9.6 fb$^{-1}$ and constitute the first search
for $X(5568)$ production in the same initial conditions as the D0
observation.  

The CDF II detector is described in detail elsewhere 
\cite{CDF_detector}.  
This analysis uses the tracking and
muon identification systems.
The tracking system measured the trajectories of charged particles (tracks)
and consisted of five layers of
double-sided silicon detectors \cite{CDF_Silicon} 
and a 96 layer open-cell drift 
chamber (COT) \cite{CDF_COT} 
that operated inside a solenoid with a 1.4 T field oriented
along the beam direction. 
Charged particles with transverse momentum $(p_T)$
greater than 250 MeV/$c$
that originated from the collision point were measured in the tracking system
with a transverse-momentum resolution of 
$\sigma(p_T)/p_T^{2} \approx 0.0008 \, $(GeV/$c)^{-1}$.
Muon candidates from the decay $J/\psi \rightarrow \mu^+ \, \mu^-$
were  identified by  two sets of drift  chambers located radially
outside   electromagnetic and   hadronic  calorimeters \cite{CDF_Muon}.
The central muon chambers
covered the pseudorapidity region $ |\eta| <  0.6$ 
and  were sensitive to muons with 
$p_T > 1.4 $ GeV/$c$.
A second muon system covered the region $  0.6 < |\eta| <  1.0$ and
detected  muons having $p_T > 2.0 $ GeV/$c$.  

The mass resolution and acceptance for the
$X(5568)$ and $B^0_s$ decays are
studied with a Monte Carlo simulation that generates 
$X(5568) \rightarrow B^0_s \, \pi^{\pm}$ decays
consistent with CDF measurements of $p_T$ and rapidity distributions
for inclusive $B^0_s$ production.
The simulated $X(5568)$ and $B^0_s$ decay isotropically and
other final-state decay processes are simulated with 
the {\sc EvtGen} \cite{EvtGen} program.
The generated events are 
passed through the detector and trigger simulation based on a 
{\sc GEANT3} description \cite{Geant}  
and processed through 
the same reconstruction and
analysis algorithms used for the data.

This analysis is based on events recorded with a 
three-level trigger
that was dedicated to the collection of a 
$J/\psi \rightarrow \mu^+ \, \mu^-$ sample.
The first level of the trigger system required two
muon candidates with tracks in the COT and muon chamber systems
that matched in the plane transverse to the beam direction.
At this stage, the trigger system
identified trigger tracks with $p_T > 1.4 $ GeV/$c$
and segmented into  $1.25\,^{\circ}$ in the azimuthal angle.
The second level imposed the requirement that the muon candidates
have opposite charge 
and limited the accepted range of opening angle
between them \cite{Bsigma}.  The third level 
of the trigger reconstructed the muon pair with the full resolution of the 
COT, and 
required that the invariant mass of the pair fall within the range
2.7--4.0 GeV/$c^2$.

The strategy for this analysis is to reconstruct the $B^0_s \, \pi^{\pm}$
final state using similar methods 
to those used by previous CDF analyses \cite{Bsigma,CDF_Baryons}.  
The measured yields and acceptances
are used to calculate the fraction of $B^0_s$ produced through
the process $X(5568) \rightarrow B^0_s \, \pi^{\pm}$, given by 
\begin{equation} 
f_{B^0_{s}/X(5568)} = \frac{\sigma(p {\bar p} \rightarrow X(5568) + x) \times
{\cal B}(X(5568)\rightarrow B^0_s \, \pi^{\pm})}
{\sigma(p {\bar p} \rightarrow B^0_s + x)} = 
\frac{N_{X}}{N_{B^0_{s}}} \frac{1}{\alpha_{X,B^0_{s}}},
\label{eq:f_XBs}
\end{equation} 
where $N_X$ and $N_{B^0_s}$ are the numbers of 
$X(5568)$ and  $B^0_s$ reconstructed in the data, respectively, and
$\alpha_{X,B^0_s}$ is the acceptance  and reconstruction efficiency
for the $X(5568)$ in events
where the $B^0_s$ is reconstructed.  In the absence of an
$X(5568)$ signal, this expression is used to calculate a limit on
$f_{B^0_s/X(5568)}$.

The analysis of the data begins with a selection of well-measured
$J/\psi \rightarrow \mu^+ \, \mu^-$ candidates.
The trigger requirements are confirmed by selecting
 events that contain two
 oppositely  charged  muon candidates, each with matching 
 COT and muon chamber tracks.
Both muon tracks are required to have associated 
measurements in at least three layers of the silicon detector,
and are fit with the constraint that they originate from a common decay point.
Dimuon candidates are
measured with an average mass resolution of 20 MeV/$c^2$ and
candidates with a mass within 80 MeV/$c^2$ of the
world-average  $J/\psi$ mass \cite{PDG} are retained as 
$J/\psi$ candidates. 
Approximately 15 million $J/\psi$ candidates are obtained.

The reconstruction of $\phi$
candidates uses all additional tracks 
found in each event containing a $J/\psi$ candidate.  
Pairs of oppositely charged tracks with 
three or more silicon layer measurements and $p_T > 400$ MeV/$c$
are assigned the $K^{\pm}$ mass
and have their
track parameters recalculated according to a fit that constrains
them to intersect.  Candidates whose fits converge have
a mass measurement resolution that is insignificant compared to the $\phi$
natural width of 4.2 MeV/$c^2$ \cite{PDG}, and those with
an invariant mass within 10 MeV/$c^2$ 
of the world-average $\phi$ mass \cite{PDG}
are retained as $\phi$ candidates.  

\begin{figure}[hbt]
\begin{center}
\psfig{figure=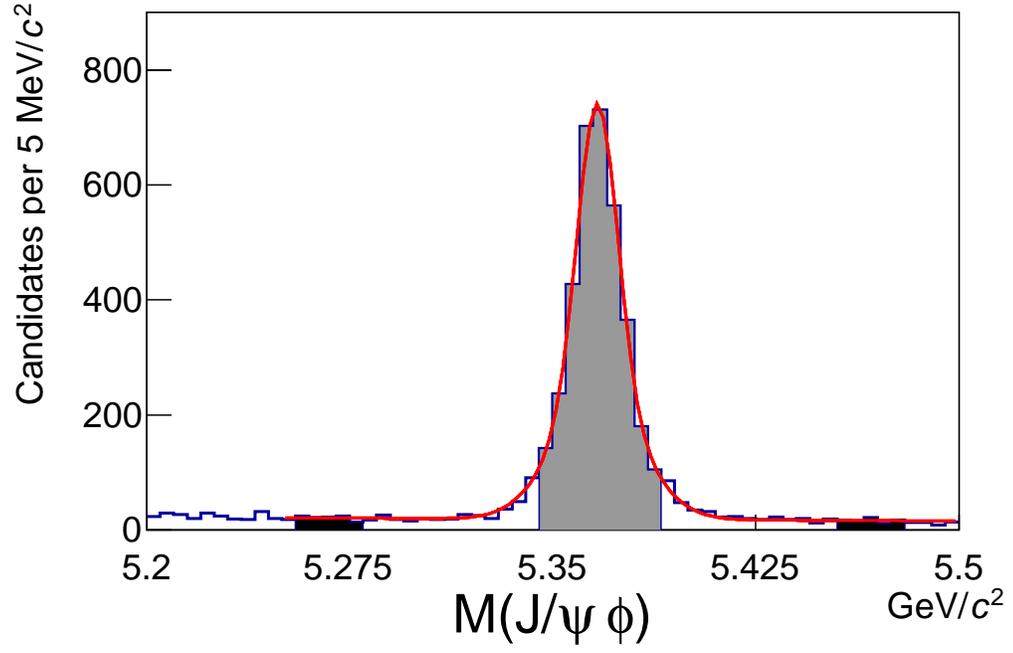,height=4.0in} 
\caption{Distribution of $J/\psi \, \phi$ mass for $p_T > 10$ GeV/$c$
with the fit overlaid on the histogram.
The $B^0_s$ signal region is highlighted in gray.  Areas used to define
backgrounds based on the mass sidebands are indicated in black.
 \label{fig:fig1}}
\end{center}
\end{figure}

The sample of $B^0_s$ candidates
is obtained by selecting all candidates where the four tracks
satisfy a fit that constrains the tracks to originate from a common 
decay point and the dimuon to have the world-average $J/\psi$ mass \cite{PDG}.  
Further requirements placed on the $B^0_s$ candidates include 
$p_T > 10 $ GeV/$c$ and $ct > 100 ~\mu$m, where $t$ is the proper
decay time of the candidate.  
These requirements 
remove candidates for which the acceptance of the detector is low
and reduce background due to prompt combinations.  
The $J/\psi \, \phi$ mass distribution obtained is shown in 
Fig.~\ref{fig:fig1}.  The number of $B^0_s$ candidates in the data is 
obtained by performing a binned likelihood fit on the distribution in
Fig.~\ref{fig:fig1} with a linear background model and two Gaussian
functions
with a common central value as a signal model.  
We measure a $B^0_s$ yield of $N_{B^0_s} \, = \, 3552 \pm 65$ candidates and a 
mass of $5366.5 \pm 0.2$ MeV/$c^2$, 
consistent with the world average \cite{PDG}.
The average mass measurement resolution is 11 MeV/$c^2$ and 
candidates with mass
within 30 MeV/$c^2$ of the nominal $B^0_s$ mass 
are used for the $X(5568)$ search.
Mass sidebands are also indicated in Fig.~\ref{fig:fig1} and are
defined as two mass ranges of 30 MeV/$c^2$ full width 
centered $\pm 100$ MeV/$c^2$
from the nominal $B^0_s$ mass.

The final $B^0_s \, \pi^{\pm}$ sample is obtained by 
combining the $B^0_s$ candidate tracks
with the remaining tracks,
assumed to be pions, 
that have three or more silicon detector hits
and $p_T > 400$ MeV/$c$.
A constrained fit is performed and requires 
the $B^0_s$ and $\pi^{\pm}$ 
candidates to originate from the same point.  This final selection 
also requires
the transverse displacement of the full final state
 with respect to the beamline to be less than $100 \, \mu$m.
A mass resolution of 1.8 MeV/$c^2$ is obtained for 
the $B^0_s \, \pi^{\pm}$ final state
by defining 
$M(B^0_s \, \pi^{\pm}) = M(J/\psi \, \phi \, \pi^{\pm}) -  M(J/\psi \, \phi) +
M_{B^0_s}$, where $M_{B^0_s}$ is the world-average value of 
the mass of the $B^0_s$ \cite{PDG}.

Simulated events are used to estimate the 
acceptance of the $B^0_s$ and the relative acceptance $\alpha_{X,B^0_{s}}$ 
for the $X(5568)$
for events containing a reconstructed $B^0_s$.
A correction is made to the generated $X(5568)$ sample
so that the simulated $p_T(B^0_s)$ distribution is identical to
the acceptance-corrected $p_T(B^0_s)$ distribution observed in the data.
Three simulated samples are generated using widths of the $X(5568)$,
with values of
21.6, 15.5, and 28.7 MeV/$c^2$.  These correspond to the central value and 
the range of the uncertainty for the width measured in the 
reported $X(5568)$ \cite{DZero}. 

The shape of the $B^0_s \, \pi^{\pm}$ mass distributions obtained from 
simulated events
is dependent on $p_T$, due to the acceptance and efficiency of 
the tracking system.  The reconstructed signal shape expected for the
$X(5568)$ is obtained by integrating the $p_T$-dependent
mass distribution shapes found in simulation with a weighting determined
by the observed $p_T(B^0_s)$ distribution.  
The expected mass-distribution 
shape is parametrized with an empirical function
using two Gaussians and a tail term on the high-mass side
from the peak. 
The yield of $X(5568)$ observed in the simulated events 
provides a value of 
$\alpha_{X,B^0_{s}} = 0.445 \pm 0.027$ for $p_T(B^0_s)~>~10$~GeV/$c$,  
where the uncertainty is statistical.  
A systematic variation on
$\alpha_{X,B^0_{s}}$ of $\pm0.018$ is found due to the uncertainty on the 
reported width of the $X(5568)$.

\begin{figure}[hbt]
\begin{center}
\psfig{figure=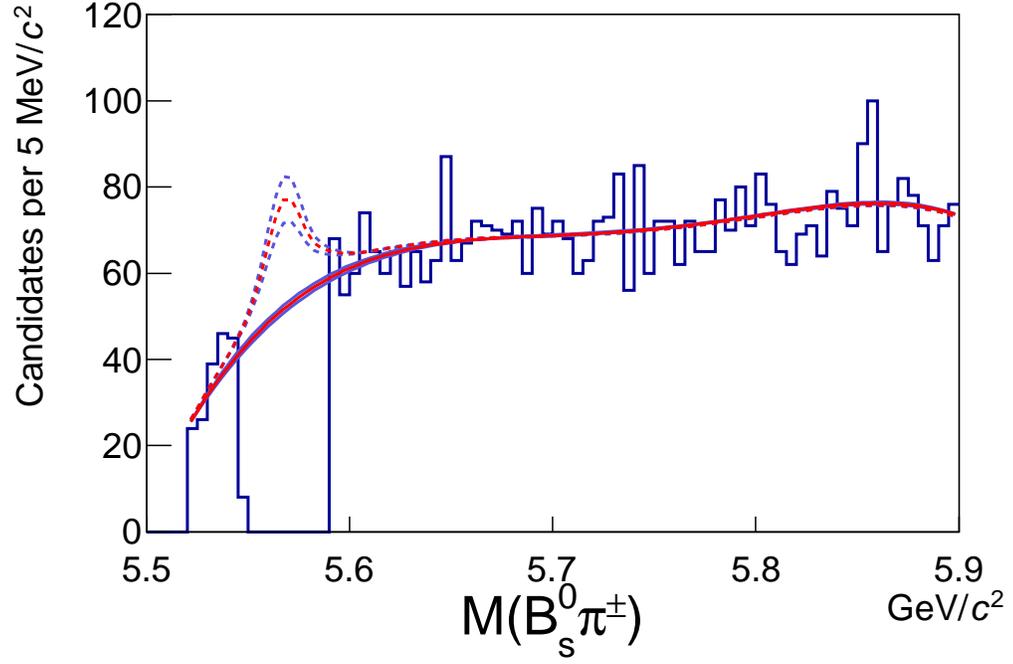,height=4.0in} 
\caption{Distribution of $B^0_s \pi^{\pm}$ mass 
where the candidates within 21.6 MeV/$c^2$ of the X(5568) are omitted.
The  background model is overlaid in a solid line,
where the line width indicates 
$\pm \sigma$ variations on the background model due to variations in 
the assumption for the $X(5568)$ signal amplitude.
Dashed curves indicate the signal components used to 
obtain the background model and its variations.
 \label{fig:fig2}}
\end{center}
\end{figure}

The $B^0_s \, \pi^{\pm}$ mass distribution
is analyzed with an unbinned likelihood fit with the likelihood calculated as
\begin{equation}
{\cal L}  =  \prod_i^{N} \left[ f{\cal S}(m_i) + 
(1 - f) {\cal B}(m_i) \right],
\label{eq:likely}
\end{equation}
where 
$N$ is the number of entries in the distribution,
$m_i$ is the mass of entry $i$,
$f$ is the signal fraction obtained from the fit,
${\cal S}(m_i)$ is the signal model obtained from simulation and
${\cal B}(m_i)$ is the background model.
The functional form of ${\cal B}(m_i)$ is obtained by fitting the
mass distribution with all candidates within the 
central value of the reported width (21.6~MeV/$c^2$) \cite{DZero}
of the $X(5568)$ mass value omitted, 
with $f$ fixed at the value
that results from the D0 observation, and ${\cal B}(m_i)$ modeled
with a polynomial.  Variations in this 
fit are also made, corresponding to the uncertainty on the
signal yield in the D0 measurement.  The background model is shown
overlaid on the data in Fig.~\ref{fig:fig2}.  

\begin{figure}[hbt]
\begin{center}
\psfig{figure=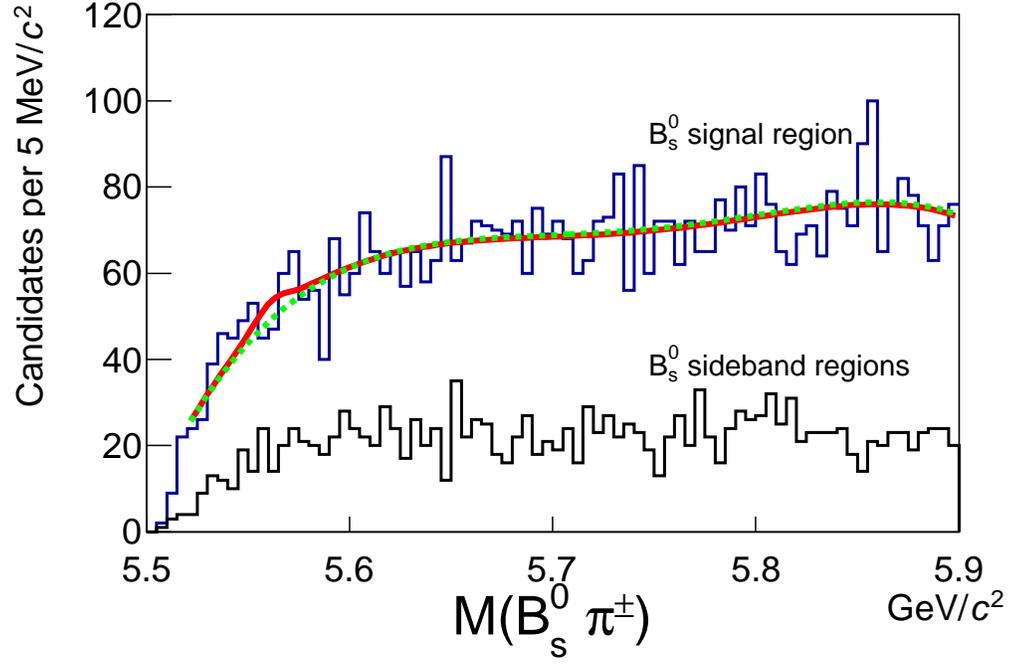,height=4.0in} 
\caption{Distribution of $B^0_s \, \pi^{\pm}$ mass for 
$p_T(B^0_s) > 10$ GeV/$c$, for candidates in the $B^0_s$
mass range and for the mass sidebands, as indicated.
The fit to the data with a freely floating (null) signal is
overlaid in a solid (dashed) line.  
\label{fig:fig3}}
\end{center}
\end{figure}

The background model obtained in this process is fixed in the fit of the
full set of candidates where $f$ is allowed to float.
This fit is overlaid on the data in Fig.~\ref{fig:fig3} and 
estimates an $X(5568)$ yield of $N_X \, = \, 36 \pm 30$ candidates. 
The signal and background models were varied by the uncertainties
in the D0 measurements on the mass, width, and production rate
of the $X(5568)$ to provide a systematic uncertainty estimate of 
14 candidates.
This signal yield is used in Eq.~(\ref{eq:f_XBs}) with the
acceptance and $B^0_s$ yield to calculate 
$f_{{B^0_s}/X(5568)} = (2.3 \pm1.9(\textrm{stat}) \pm0.9(\textrm{syst}))$\%.
Fig.~\ref{fig:fig3} also shows the
$M(B^0_s \, \pi^{\pm})$ distribution 
obtained from the $J/\psi \phi$ candidates in the $B^0_s$ 
mass sidebands indicated 
in Fig.~\ref{fig:fig1} for comparison.  

The $X(5568)$ yield obtained in the data is consistent with no signal.
Therefore, the fit where the signal is allowed to float is
compared to the null hypothesis by repeating the fit where the signal
component is omitted.  
The value of twice the difference in the logarithm of the likelihood,
 $2\delta \log{\cal L}$,
between the fits is then used as a measure of the compatibility of the 
data with the background-only hypothesis.
An upper  limit on the presence of an
$X(5568)$ signal is calculated by 
following a frequentist Neyman construction.
The technique uses simulated 
mass distributions
with a shape given by the probability distribution in Eq. (\ref{eq:likely}).
Various signal-strength hypotheses are generated
by fixing 
$f_{B^0_{s}/X(5568)}$ for each simulation, producing a signal fraction 
given by $f = N_X/N$.
Ten thousand trials are run for each
signal-strength hypothesis and the mass distributions
obtained in each trial are fit twice as in the data,
once with a floating signal 
fraction and once with the null signal hypothesis.  The $2\delta \log{\cal L}$ 
is then evaluated for all simulated distributions.

The results of these simulations are used to set
an upper limit on $f_{B^0_{s}/X(5568)}$.  The
cumulative probability distribution of $2\delta \log{\cal L}$
for a given $f_{{B^0_s}/X(5568)}$
provides the test statistic.  
The 95\% confidence level, C.L., upper limit is obtained by determining the value 
of $f_{{B^0_s}/X(5568)}$ where the value of the cumulative
probability distribution, evaluated at the value of  $2\delta \log{\cal L}$
seen in the data, approximates 0.05.
A value of $f_{{B^0_s}/X(5568)} = 0.055$ is 
obtained by this method.

Alternative background models that use a 
$B^0 \rightarrow J/\psi \, K^{*}(892)^{0}$ sample have also
been used to obtain upper limits
on $f_{{B^0_s}/X(5568)}$ through this technique.  The alternative models
are found by fitting the $B^0 \pi^-$ and $B^0 \pi^+$
mass distributions and omitting mass regions
corresponding to the $B_1(5721)^+$ or $B_2^*(5747)^+$ (for $B^0 \pi^+$).  
These background models are then used to 
repeat the simulations used in the calculations
for the upper limit.  
These alternatives give upper limits on $f_{{B^0_s}/X(5568)}$ 
that are comparable to the result based on the 
$B^0_s \, \pi^{\pm}$ fit.

Two systematic effects are considered that modify the upper limit 
obtained for $f_{{B^0_s}/X(5568)}$.  The first takes into account 
uncertainties on the quantities 
in Eq. (\ref{eq:f_XBs}).  The uncertainties on the quantities are combined in 
quadrature to obtain an overall relative uncertainty of 6.6\%, 
where the uncertainty on the
acceptance $(\alpha_{X,B^0_{s}})$ provides the largest 
contribution (6.0\%).  
The second systematic uncertainty is due to the background model. 
An alternative model, where the input value of  $f_{{B^0_s}/X(5568)}$
is increased by one standard deviation with respect to the D0 central value
tests this sensitivity.
An upper limit of $f_{{B^0_s}/X(5568)} = 0.060$ is found for
this background model, so a relative uncertainty of 9\%
is assigned to this effect.  Combining these in quadrature gives a 
total systematic uncertainty of 11\% on 
this measurement of $f_{{B^0_s}/X(5568)}$.  
We treat the systematic uncertainty as a normal distribution width, and
consider twice its value to correspond to a 95\% fluctuation.
Consequently, inclusion of the systematic uncertainty provides a
95\% C.L. upper limit on $f_{{B^0_s}/X(5568)}$ of 0.067.

In conclusion, a search for the exotic meson $X(5568)$ is performed 
with the full CDF II data set, which was obtained with the same
collision conditions and similar kinematic range as in the original
observation of this state by D0 \cite{DZero}.
No statistically significant evidence for the process
$X(5568) \rightarrow B^0_s \, \pi^{\pm}$ is found.
A 95\% C.L. upper 
limit of 6.7\% is found for  fraction of $B^0_s$ produced through this process.
Consequently, this analysis does not confirm the existence of the $X(5568)$. 

This document was prepared by the CDF collaboration using the resources
of the Fermi National Accelerator Laboratory (Fermilab), a U.S.
Department of Energy, Office of Science, HEP User Facility. Fermilab is
managed by Fermi Research Alliance, LLC (FRA), acting under Contract No.
DE-AC02-07CH11359.  We thank the Fermilab staff and the technical staffs
of the participating institutions for their vital contributions. This
work was supported by the U.S. Department of Energy and National Science
Foundation; the Italian Istituto Nazionale di Fisica Nucleare; the
Ministry of Education, Culture, Sports, Science and Technology of Japan;
the Natural Sciences and Engineering Research Council of Canada; the
National Science Council of the Republic of China; the Swiss National
Science Foundation; the A.P. Sloan Foundation; the Bundesministerium
f\"ur Bildung und Forschung, Germany; the Korean World Class University
Program, the National Research Foundation of Korea; the Science and
Technology Facilities Council and the Royal Society, United Kingdom; the
Russian Foundation for Basic Research; the Ministerio de Ciencia e
Innovaci\'{o}n, and Programa Consolider-Ingenio 2010, Spain; the Slovak
R\&D Agency; the Academy of Finland; the Australian Research Council
(ARC); and the EU community Marie Curie Fellowship Contract No. 302103.

\end{document}